\documentclass[aps,prl,twocolumn,preprintnumbers]{revtex4}
\usepackage{CJK,psfrag,graphicx,epsfig,amsmath,amssymb,bm}

\def\beq{\begin{equation}}
\def\eeq{\end{equation}}
\def\beqa{\begin{eqnarray}}
\def\eeqa{\end{eqnarray}}

\begin{document}
\title{A Twistorial Foundation for the Classical Double Copy}
\author{Chris D. White} 
\affiliation{Centre for Research in String
  Theory, School of Physics and Astronomy, Queen Mary University of
  London, 327 Mile End Road, London E1 4NS, UK} 

\preprint{QMUL-PH-20-32}

\begin{abstract}
The classical double copy relates exact solutions of gauge, gravity
and other theories. Although widely studied, its origins and domain of
applicability have remained mysterious. In this letter, we show that a
particular incarnation - the Weyl double copy - can be derived using
well-established ideas from twistor theory. As well as explaining
where the Weyl double copy comes from, the twistor formalism also
shows that it is more general than previously thought.
\end{abstract}

\maketitle




\section{Introduction}
\label{sec:intro}

The study of fundamental physics is dominated by (non-abelian) gauge
theories, which underly particle physics, and General Relativity (GR),
which describes astrophysics and cosmology. Intriguing similarities
between these theories have emerged in recent years, and we will here
concentrate on the {\it classical double copy} that provides a map
between solutions of different field equations, itself inspired by a
similar procedure for (quantum) scattering
amplitudes~\cite{Bern:2010ue,Bern:2010yg}. In general, one can only
relate classical solutions at a fixed order in perturbation
theory~\cite{Luna:2016hge,Goldberger:2016iau,Goldberger:2017frp,Goldberger:2017vcg,Goldberger:2017ogt,Shen:2018ebu,Carrillo-Gonzalez:2018pjk,Plefka:2018dpa,Plefka:2019hmz,Goldberger:2019xef,PV:2019uuv,Anastasiou:2014qba,Borsten:2015pla,Anastasiou:2016csv,Cardoso:2016ngt,Borsten:2017jpt,Anastasiou:2017taf,Anastasiou:2018rdx,LopesCardoso:2018xes,Luna:2020adi,Borsten:2020xbt,Borsten:2020zgj,Luna:2017dtq,Kosower:2018adc,Maybee:2019jus,Bautista:2019evw,Bautista:2019tdr,Cheung:2018wkq,Bern:2019crd,Bern:2019nnu,Bern:2020buy,Kalin:2019rwq,Kalin:2020mvi,Almeida:2020mrg,Godazgar:2020zbv,Chacon:2020fmr}. However,
in some cases it is possible to make statements about {\it exact}
solutions~\cite{Monteiro:2014cda,Luna:2015paa,Luna:2016due,Carrillo-Gonzalez:2017iyj,Bahjat-Abbas:2017htu,Berman:2018hwd,Bah:2019sda,CarrilloGonzalez:2019gof,Banerjee:2019saj,Ilderton:2018lsf,Monteiro:2018xev,Luna:2018dpt,Lee:2018gxc,Cho:2019ype,Kim:2019jwm,Alfonsi:2020lub,Bahjat-Abbas:2020cyb,White:2016jzc,DeSmet:2017rve,Bahjat-Abbas:2018vgo,Elor:2020nqe,Gumus:2020hbb,Keeler:2020rcv,Arkani-Hamed:2019ymq,Huang:2019cja,Alawadhi:2019urr,Moynihan:2019bor,Alawadhi:2020jrv,Easson:2020esh,Casali:2020vuy,Cristofoli:2020hnk}. In
particular, the {\it Kerr-Schild double copy} of
ref.~\cite{Monteiro:2014cda} relates certain algebraically special
spacetimes in GR to an exact solution of the gauge theory field
equations. The latter is referred to as the {\it single copy} of the
given gravity solution, and there is also a {\it zeroth copy}, which
produces a solution of so-called {\it biadjoint scalar field
  theory}. A second exact classical double copy is the recent {\it
  Weyl double copy} of ref.~\cite{Luna:2018dpt}, which we discuss
below. This relies on a well-known formalism (see
e.g. refs.~\cite{Penrose:1987uia,Penrose:1986ca,Stewart:1990uf}) which
recasts the usual tensorial field equations for electromagnetism and
GR in terms of 2-component spinors $\pi_A$ and their complex
conjugates $\pi_{A'}$. Every spacetime index of a tensor field can be
converted to a pair of spinorial indices (and vice versa) by
contracting with {\it Infeld-van-der-Waerden symbols}
$\sigma^\mu_{AA'}$. Furthermore, spinors with multiple (un)primed
indices can always be decomposed into sums of products of Levi-Civita
symbols and fully symmetric spinors. For example, the field strength
tensor of electromagnetism has the spinorial translation
\begin{align}
F_{\alpha\beta}\rightarrow F_{AA'BB'}=\phi_{AB}\epsilon_{A'B'}
+\bar{\phi}_{AB}\epsilon_{A'B'},
\label{Fmunu}
\end{align}
where the symmetric spinors $\phi_{AB}$ and $\bar{\phi}_{A'B'}$ turn
out to correspond separately to the anti-self-dual and self-dual parts
in spacetime. In GR, we will be concerned with vacuum spacetimes, for
which the Riemann tensor reduces to the {\it Weyl tensor}, with
spinorial translation
\begin{align}
C_{\alpha\beta\gamma\delta}\rightarrow 
\Psi_{ABCD}\epsilon_{A'B'}\epsilon_{C'D'}+
\bar{\Psi}_{A'B'C'D'}\epsilon_{AB}\epsilon_{CD},
\label{Cdecomp}
\end{align}
such that $\Psi_{ABCD}$ ($\bar{\Psi}_{A'B'C'D'}$) is the self-dual
(anti-self-dual) part, and the former is called the {\it Weyl
  spinor}. The vacuum electromagnetic and GR equations are special
cases of the general {\it massless free-field equation}
\begin{equation}
\nabla^{AA'}\phi_{AB\ldots E}=0,\quad \nabla^{AA'}\bar{\phi}_{A'B'\ldots E'}=0,
\label{massless}
\end{equation}
where $\phi_{AB\ldots E}$ is assumed symmetric, with $2s$ indices for
a field of spin $s$, and $\nabla^{AA'}$ is the appropriate spinorial
translation of the spacetime covariant derivative. Any symmetric
spinor factorises into a symmetrised product of one-index {\it
  principal spinors}, allowing one to classify solutions of different
theories. Electromagnetic spinors are {\it null} ({\it non-null}) if
their principal spinors are (non)-proportional. Weyl spinors with no
common principal spinors are called {\it Petrov type I}, and the
possible patterns of degeneracy $\{2,1,1\}$, $\{3,1\}$, $\{2,2\}$ and
$\{4\}$ are types II, III, D and N respectively.

Given two (possibly equal) electromagnetic field strength spinors
$\phi_{AB}$, $\tilde{\phi}_{AB}$, the Weyl double copy of
ref.~\cite{Luna:2018dpt} states that one may construct a Weyl spinor
according to the rule
\begin{equation}
\Psi_{ABCD}=\frac{1}{S}\phi_{(AB}\tilde{\phi}_{CD)},
\label{WeylDC}
\end{equation}
where $S(x)$ is a scalar field. This procedure was argued to hold for
arbitrary type D vacuum spacetimes in ref.~\cite{Luna:2018dpt}, where
the scalar $S$ could then be found in particular examples by matching
both sides of eq.~(\ref{WeylDC}). All of these solutions have the
property that they linearise the Einstein equations, so that the
derivative in eq.~(\ref{massless}) may be taken to be in flat space. 

Applications of the double copy range across many different areas of
physics, including new methods for investigating gravitational waves
and / or insights into quantum gravity (see
e.g. ref.~\cite{Bern:2019prr}), connections between gauge / gravity
theories and fluid dynamics~\cite{Keeler:2020rcv}, relations between
novel optical systems and
gravity~\cite{Barnett:2014era,Sabharwal:2019ngs}, and studies of
magnetic monopoles and topology~\cite{Alfonsi:2020lub}. But quite how
general the double copy is, and where it ultimately comes from, have
up until now remained mysterious~\footnote{For tree-level scattering
  amplitudes, the double copy has a string-theoretic
  origin~\cite{Kawai:1985xq}.}, with many open questions: (i) why is
it possible to formulate an exact double copy in position space, when
the original procedure for amplitudes~\cite{Bern:2010ue,Bern:2010yg}
is in momentum space? (ii) How can one systematically fix the scalar
function $S(x)$, and is there a well-defined procedure for the {\it
  inverse zeroth copy} that relates this to a gauge theory solution?
(iii) Can one generalise the Weyl or Kerr-Schild double copies to
curved spacetime (see
refs.~\cite{Bahjat-Abbas:2017htu,Carrillo-Gonzalez:2017iyj,Prabhu:2020avf}
for related work)? (iv) Can one generalise the Weyl double copy to
less algebraically special cases (i.e. other Petrov types)? We will be
able to answer all of these questions, by travelling to twistor space!
Twistor methods have been highly successful in the study of scattering
amplitudes (see
e.g. refs.~\cite{Mason:2013sva,ArkaniHamed:2009si,Mason:2009sa,Geyer:2014fka,Casali:2015vta}). The
present study, however, constitutes the first application to the exact
double copy of classical solutions.

\section{Twistor space and Penrose transforms}
\label{sec:twistors}

Twistor space (see
e.g. refs.~\cite{Penrose:1986ca,Huggett:1986fs,Adamo:2017qyl})
$\mathbb{T}$ corresponds to the set of solutions of the {\it twistor
  equation}
\begin{equation}
\nabla_{A'}^{(A}\Omega^{B)}=0\quad\Rightarrow\quad
\Omega^A=\omega^A-ix^{AA'}\pi_{A'},
\label{twistoreq}
\end{equation}
where the second equation gives the general solution in Minkowski
space. We may thus associate solutions of eq.~(\ref{twistoreq}) with
four-component objects (``twistors'') containing a pair of spinors:
\begin{equation}
Z^\alpha=\left(\omega^A,\pi_{A'}\right).
\label{twistor}
\end{equation}
whose Minkowski space ``location'' is defined by $\Omega^A=0$. From
eq.~(\ref{twistoreq}), this implies the {\it incidence relation}
\begin{equation}
\omega^A=ix^{AA'}\pi_{A'},
\label{incidence}
\end{equation}
which is invariant under rescalings $Z^\alpha\rightarrow \lambda
Z^\alpha$. Thus, twistors obeying eq.~(\ref{incidence}) are points in
{\it projective twistor space} $\mathbb{PT}$. A fixed point $x^\mu$ in
Minkowski space maps to a complex line in $\mathbb{PT}$, corresponding
to the {\it celestial sphere} of null directions at
$x^\mu$. Considering the conjugate equation to eq.~(\ref{twistoreq}),
we may also define {\it dual twistors} $W_\alpha$ and an inner product
$Z^\alpha W_\alpha$, which turns out to be conformally invariant. Then
the {\it Penrose transform}
\begin{equation}
\phi_{A'B'\ldots C'}(x)=\frac{1}{2\pi i}\oint_{\Gamma}\pi_{E'}d\pi^{E'}
\pi_{A'}\pi_{B'} \ldots \pi_{C'}[\rho_x f(Z^\alpha)] ,
\label{Penrose}
\end{equation}
relates holomorphic twistor functions $f(Z^\alpha)$ in $\mathbb{PT}$
(i.e. involving no non-constant dual twistors) with spacetime fields,
where $\rho_x$ denotes restriction to the celestial sphere of
spacetime point $x^{AA'}$ in $\mathbb{PT}$, and $\Gamma$ is an
arbitrary contour separating any poles. For consistency, the function
$f(Z^\alpha)$ must be homogeneous under twistor rescalings with degree
$(-n-2)$, where $n$ is the number of indices appearing on the
left-hand side.  Equation~(\ref{Penrose}) solves the spin-$n$ massless
field equation of eq.~(\ref{massless}), except for the scalar case,
which instead satisfies the conformally invariant wave equation with
Ricci scalar $R$:
\begin{equation}
\left(\Box+\frac{R}{6}\right)\phi=0.
\label{wave}
\end{equation}

\section{The Weyl double copy from twistor space}
\label{sec:Weyltwistor}

We can now state the main result of our paper, namely a derivation of
the Weyl double copy from twistor space. Consider a pair of
homegeneity -4 twistor functions $\{f^{(1,2)}_{\rm EM}\}$, which will
necessarily correspond to electromagnetic solutions
$\phi^{(1,2)}_{A'B'}$ in spacetime by eq.~(\ref{Penrose}). One may
then combine them with a homegeneity -2 function $f_{\rm scal.}$
(corresponding to a spacetime scalar $\phi$) to make a homogeneity -6
function according to
\begin{equation}
f_{\rm grav.}=\frac{f^{(1)}_{\rm EM}f^{(2)}_{\rm EM}}{f_{\rm scal.}}.
\label{Weyltwist}
\end{equation}
This leads to a (linearised) gravity solution $\phi_{A'B'C'D'}$ in
spacetime, implying that there will be some sort of spacetime
relationship between electromagnetic, scalar and gravitational
fields. Our claim is that, for suitable functions, this is precisely
the mixed Weyl double copy~\footnote{We have here focused on the
  conjugate version of eq.~(\ref{WeylDC}), so that we can use the
  Penrose transform of eq.~(\ref{Penrose}).}
\begin{equation}
\phi_{A'B'C'D'}=\frac{\phi^{(1)}_{(A'B'}\phi^{(2)}_{C'D')}}{\phi}.
\label{Weylmixed2}
\end{equation}
We have here synchronised our notation with eq.~(\ref{massless}), but
$\phi_{A'B'C'D'}$ and $\phi$ are the conjugates of the quantities
$\Psi_{ABCD}$ and $S$ appearing in eq.~(\ref{WeylDC}). In order to
determine the functions we must use in eq.~(\ref{Weyltwist}), we may
rely on the observation that if a twistor function has an $m^{\rm
  th}$-order pole, its corresponding spacetime field has an
$(n-m+1)$-fold principal spinor, where $n$ is the number of spinor
indices (see e.g. ref.~\cite{Penrose:1986ca}). We may then choose the
function $f_{\rm scal.}$ to have poles at the same locations in
twistor space as the functions $f_{\rm EM}$ and $f_{\rm grav.}$, and
we may choose the order of the poles in each case so as to reproduce
eq.~(\ref{Weylmixed2}) in spacetime. To see how this works, consider
the functions
\begin{equation}
f_{m}=\frac{\left[Q_{\alpha\beta}Z^\alpha Z^\beta\right]^{-m}}{m!}
\equiv
\frac{1}{m!}\left[\frac{N^{-1}(x)}{(\xi-\xi_1(x))(\xi-\xi_2(x))}\right]^{-m},
\label{fmdef}
\end{equation}
for some constant dual twistor $Q_{\alpha\beta}$, where $m=1$, $2$ and
$3$ for the scalar, electromagnetic and gravity cases respectively. In
the second equation, we have chosen homogeneous coordinates
$\pi_{A'}=(1,\xi)$, $\xi\in\mathbb{C}$. The resulting roots in $\xi$
and normalisation factor $N(x)$ gain their position dependence from
the use of the incidence relation of eq.~(\ref{incidence}). One then
finds that the Penrose transform in the general spin $(m-1)$ case is
\begin{align}
\underbrace{\phi_{A'B'\ldots C'}}_{(m-1)\ {\rm indices}}=\frac{N^m(x)}{2\pi i}\oint_\Gamma d\xi 
\frac{\overbrace{(1,\xi)_{A'}(1,\xi)_{B'}\ldots (1,\xi)_{C'}}^{(m-1)
\ {\rm factors}}}{(\xi-\xi_1)^m(\xi-\xi_2)^m}.
\label{Pentrans}
\end{align}
The contour $\Gamma$ is defined on the Riemann sphere of $\xi$, and is
such that it must separate the poles at $\xi=\xi_1$ and
$\xi=\xi_2$. Choosing to enclose the first of these poles, one finds
spacetime fields
\begin{align}
\phi&=\frac{N(x)}{\xi_1-\xi_2},\quad \phi_{A'B'}=-\frac{N^2(x)}{(\xi_1-\xi_2)^3}
\alpha_{(A'}\beta_{B')}\notag\\
&\quad \phi_{A'B'C'D'}=\frac{N^3(x)}{(\xi_1-\xi_2)^5}
\alpha_{(A'}\beta_{B'}\alpha_{C'}\beta_{D')},
\label{fields}
\end{align}
where the principal spinors occuring in these equations are given by
\begin{equation}
\alpha=(1,\xi_1),\quad \beta=(1,\xi_2).
\label{abdef}
\end{equation}
It is evident that the fields of eq.~(\ref{fields}) obey
eq.~(\ref{Weylmixed2}), as required. Furthermore, the gravity field in
eq.~(\ref{fields}) is clearly of type D, where the right pattern (2,2)
of degenerate principal spinors arises from choosing two distinct
third order poles for $f_{\rm grav.}$ in twistor space.

To illustrate the above, we may consider a simple special case of
eq.~(\ref{fmdef}), namely the self-dual Schwarzschild / Taub-NUT
solution, which is generated by choosing~\footnote{We have been
  inspired by a similar function for anti-self-dual fields in
  ref.~\cite{Hughston:1979tq}.}
\begin{equation}
Q_{\alpha\beta}=\frac{1}{2}\left(\begin{array}{rrrr}0&0&0&-1\\
0&0&1&0\\
0&1&0&0\\
-1&0&0&0\end{array}\right).
\label{SchwarzschildQ}
\end{equation}
One then finds
\begin{equation}
\xi_{1,2}=\frac{-z\pm r}{(x+iy)}=\frac{x-iy}{z\pm r},
\quad N(x)=\frac{i\sqrt{2}}{(x+iy)},
\label{xi12SZ}
\end{equation}
where $r=\sqrt{x^2+y^2+z^2}$ is the spherical radial coordinate. From
eq.~(\ref{fields}), the biadjoint scalar function $\phi$ associated with
this solution is given by
\begin{equation}
\phi=\frac{i}{r\sqrt{2}}.
\label{phiSZ}
\end{equation}
This agrees with the function $S(x)$ presented in
ref.~\cite{Luna:2018dpt}, up to an overall normalisation
constant. However, their function $S(x)$ is itself only defined up to
an overall constant, so this is not a problem. Furthermore, given the
principal spinors of eqs.~(\ref{abdef}, \ref{xi12SZ}), one may convert
to the Kerr-Schild form of the classical double copy by contracting
each with the relevant Infeld-van-der-Waerden
symbols~\cite{Penrose:1987uia} to obtain the spacetime vectors:
\begin{equation}
k_\mu^{\pm}\propto\left(1,\pm\frac{x}{r},\pm\frac{y}{r},
\pm\frac{z}{r}\right).
\label{kmupm}
\end{equation}
These are indeed the two possible choices of null vector entering the
Kerr-Schild double copy approach of ref.~\cite{Monteiro:2014cda}.

In general, the Weyl double copy becomes especially elegant in twistor
space: it relies on simple products of scalar functions, whereas the
spacetime fields involve products of lower-rank spinors followed by
symmetrisation over indices. However, the twistor space formulation is
much more than a simple rewriting. Once one has found the set of
functions in eq.~(\ref{fmdef}) for use in eq.~(\ref{Weyltwist}), the
known properties of the Penrose transform guarantee that there exist
corresponding spacetime fields, and that these obey the Weyl double
copy. Furthermore, the functional form of eq.~(\ref{fmdef}) is
sufficient to produce the self-dual part of the most general type D
vacuum solution~\cite{Hughston:1979tq,Haslehurst}, thus encompasses
the solutions considered in ref.~\cite{Luna:2018dpt}. Hence the
twistor framework provides a {\it derivation} of both the form, and
the previously considered scope, of the Weyl double copy. In doing so,
it also explains why the Weyl double copy (and its related Kerr-Schild
counterpart) operate directly in position space, as it is the latter
that arises from the Penrose transform.

Our argument here is confined to linearised equations of motion only,
due to the limitations of the Penrose transform. However, for the type
D vacuum solutions considered in ref.~\cite{Luna:2018dpt}, all of them
linearise their respective field equations, so that the Weyl double
copy can be promoted to an exact statement. It may also be
generalised, in principle, to arbitrary conformally flat background
spacetimes, formalising previous exploratory work in this
direction~\cite{Carrillo-Gonzalez:2017iyj,Bahjat-Abbas:2017htu}. To
see this, note that the twistor description is manifestly conformally
invariant (as stated above), so that upon obtaining the spacetime
fields of eq.~(\ref{fields}), one may transform each one to a given
background spacetime according to the usual conformal transformation
rules for multi-index spinors~\cite{Penrose:1986ca}. The resulting
fields can then be interpreted as related by the double copy in the
new background.

\section{The inverse zeroth copy}

As mentioned above, it has not previously been clear how to fix the
scalar function that appears in the Weyl double copy. In both this and
the Kerr-Schild approach, it is also not obvious how to precisely
formulate an {\it inverse zeroth copy} that relates a given scalar
field to corresponding gauge and gravity solutions. The latter contain
extra kinematic information (associated with the principal null
directions of the field strength and Weyl tensor respectively) 
that appears entirely absent in the biadjoint theory. The twistor
approach solves this problem: the principal null directions of the
gauge and gravity fields are uniquely fixed by the poles of the
corresponding twistor space functions, as evidenced directly in
eq.~(\ref{abdef}). What's more, this information is {\it already
  present} in the scalar function ($m=1$) of eq.~(\ref{fmdef}). The
twistor picture thus reveals, for the first time, how the biadjoint
field ``knows'' about the structure of the resulting gauge and gravity
fields!

\section{Beyond type D solutions}

So far, we have reproduced the type D Weyl double copy of
ref.~\cite{Luna:2018dpt}, by choosing a particular set of functions
(eq.~(\ref{fmdef})) for use in eq.~(\ref{Weyltwist}). However, we can
clearly allow for a more general set of functions to be used, and in
doing so the twistor language allows us to extend the Weyl double copy
to solutions other than Petrov type D. 

To find a concrete example, we may use a particularly well-studied
class of holomorphic twistor functions, namely {\it elementary states}
(see e.g. ref.~\cite{Penrose:1986ca}), which consist of ratios of
factors of the form $(A_\alpha Z^\alpha)$, where $A_\alpha$ is a
constant dual twistor. Such functions were originally motivated as
alternatives to plane-wave states in examining scattering processes
via twistor space, but have been reconsidered in a recent series of
papers~\cite{Dalhuisen:2012zz,Swearngin:2013sks,deKlerk:2017qvq,Sabharwal:2019ngs,Thompson:2014owa,Thompson:2014pta},
where they are shown to give rise to topologically non-trivial
configurations of electromagnetic and gravitational fields, where the
field lines form {\it torus knots}. Knotted magnetic fields are of
great interest due to their potential role in stabilising nuclear
fusion processes, and for stellar
structure~\cite{Arrayas:2017sfq}. Furthermore, finding gravitational
counterparts of interesting electromagnetic solutions may guide
experimental efforts to emulate gravitational
waves~\cite{Fernandez-Corbaton:2014cha}.

In particular, consider the
Penrose transform pair~\cite{Thompson:2014owa}
\begin{align}
&\frac{1}{(A_\alpha Z^\alpha)^{1+a}(B_\alpha Z^\alpha)^{1+b}}
\rightarrow\\
&\quad \left(\frac{2}{\Omega|x-y|^2}\right)^{a+b+1}
{\cal A}_{(A'_1}\ldots{\cal A}_{A'_b}{\cal B}_{A'_{b+1}}\ldots
{\cal B}_{A'_{2h})},
\label{knots1}
\end{align}
where $a,b\in\mathbb{Z}$, the curly spinors are defined by
\begin{equation}
A_\alpha Z^\alpha\equiv {\cal A}^{A'}\pi_{A'},\quad
B_\beta Z^\beta\equiv {\cal B}^{B'}\pi_{B'},
\label{ABcaldef}
\end{equation}
and
\begin{align}
A_\alpha&=(\mu_A,\lambda^{A'}),\quad B_\alpha=(\sigma_A,\psi^{A'}),\notag\\
\Omega=&\mu_B\sigma^B,\quad y^{AA'}=i\frac{\sigma^A\lambda^{A'}-\mu^A\psi^{A'}}
{\mu_B\sigma^B}.
\label{Omegadef}
\end{align}
For $a=b=0$, we obtain the scalar field
\begin{equation}
\phi=\frac{2}{\Omega|x-y|^2}.
\label{hopfscalar}
\end{equation}
One may construct twistor functions of
homogeneity -4 by choosing $(a,b)=(1,1)$ or $(0,2)$, leading to the two
respective electromagnetic spinors
\begin{align}
\phi^{(1,1)}_{A'B'}&=\left(\frac{2}{\Omega|x-y|^2}\right)^3{\cal A}_{A'}
{\cal B}_{B'},\notag\\
\phi^{(0,2)}_{A'B'}&=\left(\frac{2}{\Omega|x-y|^2}\right)^3{\cal A}_{A'}
{\cal A}_{B'}.
\label{emspinors}
\end{align}
Using these in the mixed Weyl double copy of eq.~(\ref{Weylmixed2}),
one can generate a number of different Weyl spinors:
\begin{align}
\phi^{(1,1)\times(1,1)}_{A'B'C'D'}&=\left(\frac{2}
{\Omega|x-y|^2}\right)^5{\cal A}_{(A'}{\cal A}_{B'}{\cal B}_{C'}
{\cal B}_{D')},\notag\\
\phi^{(1,1)\times (0,2)}_{A'B'C'D'}&=\left(\frac{2}
{\Omega|x-y|^2}\right)^5{\cal A}_{(A'}{\cal A}_{B'}{\cal A}_{C'}
{\cal B}_{D')},\\
\phi^{(0,2)\times (0,2)}_{A'B'C'D'}&=\left(\frac{2}
{\Omega|x-y|^2}\right)^5{\cal A}_{(A'}{\cal A}_{B'}{\cal A}_{C'}
{\cal A}_{D')},
\label{weylchoices}
\end{align}
and it is easily checked that these are the fields that arise upon
multiplying the corresponding functions in twistor space according to
eq.~(\ref{Weyltwist}), and then performing the Penrose transform to
position space. The first and third of these examples are Petrov type
D and N respectively. However, the second (as already noted in
ref.~\cite{Thompson:2014owa}) is Petrov type III. This thus goes
beyond the original formulation of the Weyl double copy in
ref.~\cite{Luna:2018dpt}. The price one pays, however, is that such
solutions are restricted to linear level only.

\section{Conclusion}
\label{sec:conclude}

We have presented a twistor space derivation of the exact classical
double copy, that reproduces the Weyl double copy of
ref.~\cite{Luna:2018dpt}, itself equivalent to the Kerr-Schild double
copy of ref.~\cite{Monteiro:2014cda}, where they overlap. It resolves
a number of questions, which we labelled above: (i) the Penrose
transform of eq.~(\ref{Penrose}) relates twistor functions to
spacetime fields in {\it position space}, thus explaining why a
position space exact copy is possible; (ii) the scalar function $S(x)$
in eq.~(\ref{WeylDC}) is predicted exactly by the twistor approach,
and its poles in twistor space already ``know'' what the principal
spinors in the gauge and gravity solutions will be, thus providing an
explicit interpretation for the inverse zeroth copy; (iii) conformal
invariance of the twistor space formulation implies that the classical
double copy should immediately generalise to conformally flat
spacetimes, formalising the exploratory results of
refs.~\cite{Bahjat-Abbas:2017htu,Carrillo-Gonzalez:2017iyj}, and where
the biadjoint field should obey eq.~(\ref{wave}); (iv) our new
approach leads to Petrov types other than type D (or N), thereby
broadening the scope of the Weyl double copy. 

We expect that the twistor language could have a number of uses,
including providing new explicit examples of the classical double
copy, and to ascertain its scope (e.g. by showing which Petrov types
are (not) possible). In line with the general remarks above, it would
be interesting to formulate explicit examples of double copies in
conformally flat backgrounds, including those of astrophysical
relevance. We also note that the twistor language can in principle be
extended beyond linear level, using appropriate generalisations of
the Penrose transform~\cite{Ward:1977ta,Penrose:1976jq}.

Our methods have been manifestly four-dimensional. For higher
dimensions, it may be more sensible to use an {\it ambitwistor}
approach in which (dual) twistors are placed on a more equal footing,
as has proven useful for scattering
amplitudes~\cite{Mason:2013sva,ArkaniHamed:2009si,Mason:2009sa,Geyer:2014fka,Casali:2015vta}. In any case,
given the role that twistor theory has played in many different areas
of physics and mathematics~\cite{Atiyah:2017erd}, we hope that this
letter attracts the interest of communities who have been previously
unaware of the fascinating subject of the double copy.

\section*{Acknowledgments}

I thank the participants of the ``QCD Meets Gravity VI'' workshop for
useful discussions, and am especially grateful to Dr. Michael Simmonds
for his encouragement to make these results more widely available in
the first place. This work has been supported by the UK Science and
Technology Facilities Council (STFC) Consolidated Grant ST/P000754/1
``String theory, gauge theory and duality'', and by the European Union
Horizon 2020 research and innovation programme under the Marie
Sk\l{}odowska-Curie grant agreement No. 764850 ``SAGEX''.


\bibliography{refs}
  
\end{document}